\documentclass[conference]{IEEEtran}
\IEEEoverridecommandlockouts

\usepackage{cite}
\usepackage{amsmath,amssymb,amsfonts}
\usepackage{algorithmic}
\usepackage{graphicx}
\usepackage{textcomp}
\usepackage{xcolor}
\usepackage{hyperref}
\def\BibTeX{{\rm B\kern-.05em{\sc i\kern-.025em b}\kern-.08em
    T\kern-.1667em\lower.7ex\hbox{E}\kern-.125emX}}

\usepackage[usestackEOL]{stackengine}
\usepackage{import}
\usepackage{enumitem}
\usepackage{booktabs}
\usepackage{styles/cryptocode}
\usepackage[absolute]{textpos}
\newcommand\submittedtext{%
\footnotesize \textbf{Preprint}.%
This work has been accepted to the %
IEEE International Workshop on Programmable Zero-Knowledge Proofs for Decentralized Applications (ZKDAPPS 2025), %
in conjunction with the IEEE International Conference on Blockchain and Cryptocurrency (ICBC 2025). © 2025 IEEE. Personal use of this material is permitted. Permission from IEEE must be obtained for all other uses, in any current or future media. }

\newcommand\submittednotice{%
\begin{tikzpicture}[remember picture,overlay]
\node[anchor=south,yshift=10pt] at (current page.south) {\fbox{\parbox{\dimexpr0.65\textwidth-\fboxsep-\fboxrule\relax}{\submittedtext}}};
\end{tikzpicture}%
}

\begin{document}
\title{From Paper Trails to Trust on Tracks: Adding Public Transparency to Railways via zk-SNARKs}

\newcommand{\scalecontent}[2]{
    \scalebox{#1}{
            \hspace{-.8cm}
        \begin{minipage}{10cm}
#2
        \end{minipage}
    }
}

\newcommand{\gls}[1]{#1}
\newcommand{\sekt}{Sektorleitlinie~22}

\newcommand{\keygen}{\ensuremath{\mathsf{KGen}}}
\newcommand{\keygeninput}{\ensuremath{1^\lambda}}
\newcommand{\keypair}{\ensuremath{sk, pk}}
\newcommand{\sk}{\ensuremath{sk}}
\newcommand{\pk}{\ensuremath{pk}}
\newcommand{\sig}{\ensuremath{\sigma}}
\newcommand{\rolesig}{\ensuremath{\sig^{\mathsf{role}}}}
\newcommand{\docsig}{\sig^{\mathsf{doc}}}
\newcommand{\action}{\ensuremath{action}}
\newcommand\variablename[1]{\mathop{\mathit{#1}}\nolimits}
\newcommand{\docinfo}{\ensuremath{\variablename{dinfo}}}
\newcommand{\roleinfo}{\ensuremath{\variablename{rinfo}}}

\newcommand{\attestdoc}{\ensuremath{\mathsf{AttestDoc}}}
\newcommand{\attestdocinput}{\ensuremath{\sk, \docinfo}}
\newcommand{\attestdocoutput}{\ensuremath{\docsig}}

\newcommand{\attestrole}{\ensuremath{\mathsf{AttestRole}}}
\newcommand{\attestroleinput}{\ensuremath{\sk, \pk, \roleinfo}}
\newcommand{\attestroleoutput}{\ensuremath{\rolesig}}

\newcommand{\verifychain}{\ensuremath{\mathsf{VerifyChain}}}
\newcommand{\verifychaininput}{\ensuremath{\docinfo, \pk_0, \roleinfo_0, \docsig, \{(\pk_i, \roleinfo_i, \rolesig_i)\}^n}}
\newcommand{\verifychainoutput}{\ensuremath{\{0, 1\}}}

\newcommand{\sign}{\ensuremath{\mathsf{Sign}}}
\newcommand{\verify}{\ensuremath{\mathsf{Verify}}}
\newcommand{\msg}{\ensuremath{\mathsf{m}}}

\newcommand{\sigscheme}{\ensuremath{\mathsf{Sig}}}
\newcommand{\sigsig}{\ensuremath{\sigma}}
\newcommand{\sigmsg}{\ensuremath{m}}
\newcommand{\sigkeygen}{\ensuremath{\sigscheme.\mathsf{KGen}}}
\newcommand{\sigsk}{\ensuremath{sk}}
\newcommand{\sigpk}{\ensuremath{pk}}
\newcommand{\sigkeypair}{\ensuremath{\sigsk, \sigpk}}
\newcommand{\sigsign}{\ensuremath{\sigscheme.\mathsf{Sign}}}
\newcommand{\sigsigninput}{\ensuremath{\sk, \sigmsg}}
\newcommand{\sigsignoutput}{\ensuremath{\sigsig}}
\newcommand{\sigverify}{\ensuremath{\sigscheme.\mathsf{Verify}}}

\newcommand{\return}{\textbf{return }}
\newcommand{\returnif}[3]{\return #1 \textbf{ if } #2 \textbf{ else } #3 }

\newcommand{\rootattestor}{\ensuremath{\mathcal{A}_r}}
\newcommand{\attestor}{\ensuremath{\mathcal{A}}}
\newcommand{\prover}{\ensuremath{\mathcal{P}}}
\newcommand{\verifier}{\ensuremath{\mathcal{V}}}

\newcommand{\chainpolicy}{\ensuremath{f_c}}
\newcommand{\phasepolicy}{\ensuremath{f_p}}
\newcommand{\chainsize}{\ensuremath{n}}
\newcommand{\treeheight}{\chainsize}
\newcommand{\ndocs}{\ensuremath{l}}

\newcommand{\rpk}{\ensuremath{rpk}}


\newcommand{\sktdoc}[1]{\ensuremath{\mathtt{#1}}}
\newcommand{\sktSA}{\sktdoc{SA}}
\newcommand{\sktTR}{\sktdoc{TR}}
\newcommand{\sktPTDD}{\sktdoc{PTD1}}
\newcommand{\sktRS}{\sktdoc{RS}}
\newcommand{\sktPTD}{\sktdoc{PTD}}
\newcommand{\sktSUC}{\sktdoc{SUC}}
\newcommand{\sktSUCC}{\sktdoc{SUC2}}
\newcommand{\sktFGV}{\sktdoc{FGV1}}
\newcommand{\sktFGVV}{\sktdoc{FGV2}}

\newcommand{\zksnark}{\ensuremath{\mathsf{ZKS}}}
\newcommand{\zksnarksetup}{\ensuremath{\mathsf{Setup}}}

\newcommand{\zksnarkprove}{\ensuremath{\mathsf{Prove}}}
\newcommand{\zksnarkproveinput}{\ensuremath{
    \zksnarkecs, \zksnarkpublicin, \zksnarkprivatein, \zksnarkwitness,
    \zksnarkprovingkey
    }}

    \newcommand{\zksnarkverify}{\ensuremath{\mathsf{Verify}}}
    \newcommand{\zksnarkverifyinput}{\ensuremath{
        \zksnarkproof, \zksnarkpublicin,
        \zksnarkverificationkey
        }}
        \newcommand{\zksnarkproof}{\ensuremath{\pi}}
        \newcommand{\zksnarkecs}{\ensuremath{ecs}}
        \newcommand{\zksnarksrs}{\ensuremath{srs}}
        \newcommand{\zksnarkcrs}{\ensuremath{crs}}
        \newcommand{\zksnarkprovingkey}{\ensuremath{zpk}}
        \newcommand{\zksnarkverificationkey}{\ensuremath{zvk}}
        \newcommand{\zksnarkpublicin}{\ensuremath{x}}
        \newcommand{\zksnarkprivatein}{\ensuremath{x'}}
        \newcommand{\zksnarkwitness}{\ensuremath{w}}

\author{
\IEEEauthorblockN{
    Tarek Galal
}
\IEEEauthorblockA{
\textit{
    Information Systems Engineering
} \\
\textit{
    TU Berlin
}\\
    Berlin, Germany
\\
    t.galal@tu-berlin.de
}

\and
    
    \IEEEauthorblockN{Valeria Tisch}
\IEEEauthorblockA{
\textit{
    Hasso Plattner Institute, 
}\\
\textit{
    University of Potsdam
}\\
Potsdam, Germany
\\
valeria.tisch@student.hpi.uni-potsdam.de}

\and

\IEEEauthorblockN{
    Katja Assaf, Andreas Polze
}
\IEEEauthorblockA{
\textit{
    Hasso Plattner Institute, 
}\\
\textit{
    University of Potsdam
}\\
    Potsdam, Germany
\\
    \{katja.assaf, andreas.polze\}@hpi.de
}

}
\maketitle
\begin{abstract}
    Railways provide a critical service and operate under strict regulatory frameworks for implementing changes or upgrades. Despite their impact on the public, these frameworks do not define means or mechanisms for transparency towards the public, leading to reduced trust and complex tracking processes.

We analyse the German guideline for railway-infrastructural modifications from proposal to approval, using the guideline as a motivating example for modelling decisions in processes using digital signatures and zero-knowledge proofs. Therein, a verifier can verify that a process was executed correctly by the involved parties and according to specification without learning confidential information such as trade secrets or identities of the participants.
We validate our system by applying it to the railway process, demonstrating how it realises various rules, and we evaluate its scalability with increased process complexities. Our solution is not railway-specific but also applicable to other contexts, helping leverage zero-knowledge proofs for public transparency and trust.

\end{abstract}


\begin{textblock*}{27cm}(0pt,280pt)
\submittednotice{}

\end{textblock*}
\section{Introduction}

Railway systems follow lengthy safety assurance processes for assessing proposals for infrastructural changes.
These processes are carried out according to established specifications that define all requirements and procedures for implementing a particular type of project.
Additionally, several national and international regulatory and accreditation
bodies may review and certify each step to ensure compliance.
Such measures are necessary for the safe operation of individual components as
well as the entire system.

Despite their direct impact on the public, these processes typically do not
have public transparency as a goal and are hard to trace.
The interested public often has to work through press releases to find information or make use of their freedom of information (FOI) rights and explicitly request documents from government bodies.
In most cases, these projects are partially or fully publicly
funded. In Germany, for example, the government is responsible for the expansion and maintenance of the railway network\footnote{DB Capital Expenditures, https://ir.deutschebahn.com/en/db-group/capital-expenditures/ (accessed 16.03.2025)}.

\paragraph*{Public Transparency Framework}
Following the idea of ``Public Money, Public Information'' that advocates for freedom of information and transparency, we believe that modifications to publicly critical infrastructures should be publicly transparent and traceable.
Therefore, we analyse the German guidelines for railway-infrastructural
modifications, known as \sekt{}, to understand typical
workflows in railway processes.
Accordingly, we propose our framework \emph{Trust on Tracks} that augments not only railway industry processes but any similar business process with public transparency without requiring modifications to the process itself.
Our framework enables modelling responsibilities and actions in a business process like railway modifications and constructs publicly verifiable zero-knowledge proofs that attest to the correctness of the process execution without revealing any potentially process-sensitive details like the identities of involved parties.
\paragraph*{Specialized Business Processes}
A business process is a structured set of activities designed to achieve a
specific organisational goal. Although the definition is not unanimous and unspecific, it is widely used \cite{MP00}.
While Business Process Modelling (BPM) helps optimise complex workflows using standardised methods and notations, we focus on a specialised solution for the category of processes similar to \sekt{}, described in \cite{sek22v21}\footnote{A newer version adding a security process became available in 2024.}.
Therein, a process is simply a hierarchy of roles with different
responsibilities and tasks.
Instead of following BPM patterns, which can add unnecessary complexity, our
specialised approach prioritises simplicity and efficiency.

\subsection{Contributions}

We present \emph{Trust on Tracks} as our framework for modelling hierarchical responsibilities in a business process and adding public transparency to the process outcomes.
The framework supports processes involving multiple participants with
hierarchical relationships amongst them, as well as specifying arbitrary
business rules for the tasks they perform.

We construct the framework using zk-SNARKs and digital signatures.
This ensures the authenticity of participants and their tasks in a process but also enables verification of arbitrary business requirements for relationships between participants or tasks and one another.
Furthermore, the zero-knowledge property of the SNARK enables confidentiality of any aspect of the process if necessary.
Finally, we use the framework to implement the first phase of the railway
specification, modelling the hierarchy of responsibilities and
business rules, and demonstrating the feasibility and practicality of the
framework.

\subsection{Related Work}

Research works investigating block\-chain applications in the railway sector
are primarily concerned with general applicability \cite{Nas18}, decentralized railway control \cite{KKJ20}, communication bandwidth sharing \cite{QLWS21}, storing and sharing information about the railway system \cite{Mun21,HZWS20,RBM+22,NZ23}, or key respectively identity management \cite{ZLS+22,FZS+23}.
In 2020, the industry seemed fond of applying blockchain, stating,
"DB\footnote{DB is the German railway operator.} is currently working on around
20 use cases for blockchain technology, including logistics supply chains, more
convenient ticketing across modes of transport, and rail
operations."\footnote{https://deutschebahn.com/en/blockchain-6935084,
(accessed 16.02.2025)} However, links to the advertised \emph{blockchain team}'s
website have been taken down. Besides such statements, blockchain technology is
not reported as being used by railway systems.

Beyond railways, several works address the use of the blockchain for adding transparency to
various business processes.
Some propose frameworks for executing business processes entirely
on-chain~\cite{weber2016untrusted,garcia2017optimized,nawari2019blockchain},
while others propose specialized solutions for concrete
sectors like supply
chain~\cite{chang2019supply}.
Here, smart contracts on the blockchain coordinate the execution of processes among different parties, allowing the processes to inherit public transparency of the blockchain.
Our approach for adding transparency to business processes does not require modelling
or executing the process on-chain.
We let the process execute off-chain, and only use the
blockchain for publishing proofs of the correctness of the process execution.

Combining zero-knowledge proofs with blockchains for transparency in business process was addressed in~\cite{toots2020zero, petto2024interpreted}.
These provide powerful frameworks for modelling entire business processes using the blockchain and zk-SNARKs, building upon Business Process Model and Notation (BPMN).
However, as we are primarily interested in the subset of processes similar to \sekt{}, we forgo the complexities of formal business process modelling and instead provide a specialized but simpler framework.

\begin{figure*}[ht]
    \centering
    \tiny

\begin{minipage}{.3\linewidth}

    \begin{tikzpicture}[scale=0.8]

        \draw[draw=white] (0,3) rectangle (12,-1);
    
        \node (GGov) at (1,3) {German Government};

        \node (FRA) at (-1,2) {FRA};
        \node (NoBo) at (3,2) {NoBo};
    
        \node (PSV) at (-2,1) {PSV};
        \node (Ops) at (-1,1) {Operator};
        \node (AsBo) at (0,1) {AsBo};
        \node (DeBo) at (1,1) {DeBo};
    
        \node (FGV) at (-2,0) {FGV};
        \node (BAV) at (-1,0) {BAV};
        
        \draw (FRA) -- (PSV);
        \draw (FRA) -- (Ops);
        \draw (FRA) -- (AsBo);
        \draw (FRA) -- (DeBo);
        
        \draw (Ops) -- (FGV);
        \draw (Ops) -- (BAV);
    
        \draw (GGov) -- (NoBo);
        \draw (GGov) -- (FRA);
    
        \draw[draw=blue] (-2.5,2.5) rectangle (1.5,-0.5);
        \node[below,blue] at (-1,-0.5){National Regulation};
    
        \definecolor{green}{RGB}{50,120,50}
        \draw[draw=green] (2,2.5) rectangle (4,1.5);
        \node[below,green] at (3,1.5){European Regulation};
    \end{tikzpicture}

    \caption{Tree structure of roles}
    \label{fig:roletree}

\end{minipage}
\hspace{0.5cm}
\begin{minipage}{.3\linewidth}

    \centering
        \begin{tikzpicture}[scale=0.4]

        \draw[draw=white] (0,8) rectangle (12,0);

        \node[orange, align=center] (FRA) at (6,9) {FRA};
    
        \node[orange, align=center] (AsBo) at (0,7.5) {AsBo};
        \node[orange, align=center] (DeBo) at (2,7.5) {DeBo};
        \node[orange, align=center] (Ops) at (6,7.5) {Operator};
    
        \node[orange, align=center] (FGV1) at (4,6) {$\text{FGV}_1$};
        \node[orange, align=center] (BAV) at (6,6) {BAV};
        \node[orange, align=center] (FGV2) at (8,6) {$\text{FGV}_2$};
    
        \node[blue, align=left] (DocSafe) at (0,4) {Doc\\SA};
        \node[blue, align=left] (DocTec) at (2,4) {Doc\\TR};
        \node[blue, align=left] (DocRep1) at (4,4) {Doc\\PTD1};
        \node[blue, align=left] (DocRS) at (6,4) {Doc\\RS};
        \node[blue, align=left] (NoteSUC2) at (9,4) {Note\\SUC2};
        
        \node[blue, align=left] (DocRep) at (3,2) {Doc\\PTD};
        \node[blue, align=left] (NoteSUC) at (8,2) {Note \\ SUC};
        \node[blue, align=left] (NoteFGV1) at (10,2) {Note \\ FGV1};
        \node[blue, align=left] (NoteFGV2) at (12,2) {Note \\ FGV2};
        
        \node[black, align=center] (LHPhase) at (7,0) {RS Phase};
    
        \draw (LHPhase) -- (DocRep);
        \draw (LHPhase) -- (NoteSUC);
        \draw (LHPhase) -- (NoteFGV1);
        \draw (LHPhase) -- (NoteFGV2);
    
        \draw (DocRep) -- (DocSafe);
        \draw (DocRep) -- (DocTec);
        \draw (DocRep) -- (DocRep1);
        \draw (DocRep) -- (DocRS);
        
        \draw (NoteSUC) -- (NoteSUC2);
    
        \draw (FRA) -- (AsBo);
        \draw (FRA) -- (DeBo);
        \draw (FRA) -- (Ops);
    
        \draw (Ops) -- (FGV1);
        \draw (Ops) -- (BAV);
        \draw (Ops) -- (FGV2);
    
        \draw[red] (AsBo) -- (DocSafe);
        \draw[red] (DeBo) -- (DocTec);
    
        \draw[red] (FGV1) -- (DocRep1);
        \draw[red] (BAV) -- (DocRS);
        \draw[red] (BAV) to[out=-50,in=90] (NoteSUC);
        \draw[red] (FGV2) to[out=-90,in=0] (DocRep);
        \draw[red] (FGV2) -- (NoteSUC2);
    
        \draw[red] (FRA) to[out=-20,in=90] (NoteFGV1);
        \draw[red] (FRA) to[out=0,in=90] (NoteFGV2);
        
    \end{tikzpicture}

        \caption{Mycorrhizal network of the requirement specification phase}
        \label{fig:RSphaseMyccNet}

    \end{minipage}
    \hspace{0.5cm}
    \begin{minipage}{.3\linewidth}

    \begin{tikzpicture}[scale=0.4]

        \draw[draw=white] (0,8) rectangle (12,-1.5);
    
        \node[orange, align=center] (FRA) at (6,7) {FRA};
    
        \node[orange, align=center] (AsBo) at (1.5,5.5) {AsBo};
        \node[orange, align=center] (DeBo) at (3,5.5) {DeBo};
        \node[orange, align=center] (Ops) at (6,5.5) {Operator};
    
        \node[orange, align=center] (FGV1) at (4,4) {$\text{FGV}_1$};
        \node[orange, align=center] (BAV) at (6,4) {BAV};
        \node[orange, align=center] (FGV2) at (8,4) {$\text{FGV}_2$};
    
        \node[blue, align=left] (DocSafe) at (1.5,3) {SA};
        \node[blue, align=left] (DocTec) at (3,2.5) {TR};
        \node[blue, align=left] (DocRep1) at (4,3) {PTD1};
        \node[blue, align=left] (DocRS) at (5,2.5) {RS};
        \node[blue, align=left] (NoteSUC2) at (8.5,3) {SUC2};
        
        \node[blue, align=left] (DocRep) at (7,2.5) {PTD};
        \node[blue, align=left] (NoteSUC) at (6,3) {SUC};
        \node[blue, align=left] (NoteFGV1) at (10,2.5) {FGV1};
        \node[blue, align=left] (NoteFGV2) at (11,3) {FGV2};

        \draw (FRA) -- (AsBo);
        \draw (FRA) -- (DeBo);
        \draw (FRA) -- (Ops);
    
        \draw (Ops) -- (FGV1);
        \draw (Ops) -- (BAV);
        \draw (Ops) -- (FGV2);
    
        \draw (AsBo) -- (DocSafe);
        \draw (DeBo) -- (DocTec);
    
        \draw (FGV1) -- (DocRep1);
        \draw (BAV) to[out=-130,in=90] (DocRS);
        \draw (BAV) -- (NoteSUC);
        \draw (FGV2) to[out=-130,in=90] (DocRep);
        \draw (FGV2) -- (NoteSUC2);
    
        \draw (FRA) to[out=-20,in=90] (NoteFGV1);
        \draw (FRA) to[out=0,in=90] (NoteFGV2);
    
        \definecolor{darkblue}{RGB}{50,50,120}
        
        \draw[->, darkblue, thick] (NoteSUC) to[out=-90,in=-90] (NoteSUC2);
    
        \draw[->, darkblue, thick] (DocRep) to[out=-70,in=-90] (DocSafe);
        \draw[->, darkblue, thick] (DocRep) to[out=-90,in=-90] (DocTec);
        \draw[->, darkblue, thick] (DocRep) to[out=-110,in=-90] (DocRep1);
        \draw[->, darkblue, thick] (DocRep) to[out=-130,in=-90] (DocRS);
    
        \draw[draw=darkblue] (0.5,3.5) rectangle (12,0.5);
        \node[below,darkblue] at (2.5,0.5){Phase Policy};
        
    \end{tikzpicture}

        \caption{Tree structure of the requirement specification phase}
        \label{fig:RSphaseTree}
    \end{minipage}

\end{figure*}

\section{Exemplary Process from the German Railway}
\label{section:railway}

The Sektorleitlinie 22 \cite{sek22v21} describes the verification process for railway applications in Germany. It consists of three main phases: \emph{requirement specification (Lastenheft)}, \emph{technical specification (Pflichtenheft)}\footnote{The most common translation for both terms is \emph{specification}.}, and \emph{product (Produkt)}, and is in the jurisdiction of three main bodies: the \emph{federal railway authority (FRA, dt. Eisenbahnbundesamt)}, the \emph{German railway operator (Deutsche Bahn)} and the vendor.
Each phase ends with a milestone, where a collection of documents, notifications, and any additional information the phase requires are submitted.

Participants involved in the process are organized in a hierarchical structure, illustrated in Figure \ref{fig:roletree}, with the German government at the root of the hierarchy, being the official primary regulator of such processes.
The \emph{notified body} (NoBo) assesses conformity with European regulations to ensure interoperability.
The \emph{common safety method assessment body} (AsBo) ensures that the system is safe.
The \emph{designated body} (DeBo) is responsible for conformity with technical regulations.
The product manager (BAV), appointed by the railway operator, is responsible for writing the requirement specification. 
Further, the railway operator may assign employees with special expertise to the approver role (FGV), and is required to notify the FRA of this assignment.

\paragraph*{Phase requirement specification (RS)}
For accessibility of our model, we assume that a railway operator initiated the development of a new product and started the requirement specification phase. We assume our new product is type B\footnote{The list of type A products is provided in the appendix of \cite{sek22v21}.} with no international interoperability requirements and two involved FGVs.
The main deliverable of the requirement specification phase is the requirement specification (DocSR), a document provided to a vendor with supplementary material, such as safety assessment (DocSA), a report on conformity with technical regulations (DocTR) and partial test declaration (DocPTD), at the end of the requirement specification phase. 

For our exemplary application scenario, Figure~\ref{fig:RSphaseMyccNet} illustrates how the deliverables in the specification requirement phase are connected.
The role tree (from above) represents the hierarchical structure of the involved organizations and roles, and the document tree (from below) represents the project phase structure and its deliverables.

The RS Phase requires three notifications to be sent to the FRA: the system to be developed (SUC), the appointment of $FGV_1$, and the appointment of $FGV_2$. $ FGV_2$ creates the final partial test declaration PTD. The declaration is supported by the safety assessment provided by the AsBo, the technical conformity report provided by the DeBo, the partial test declaration created by the other $FGV_1$ and the requirements specification itself created by the BAV.

\paragraph*{Data Structure and Graph Transformation}
The process in Figure~\ref{fig:RSphaseMyccNet} is represented by two trees.
These are connected by edges representing roles acting on documents and
notifications. Based on biology terminology where a fungal network connects
trees, we call this data structure a \emph{mycorrhizal network}.

To verify that a process phase was completed, all of its child nodes representing documents must be verifiable
such that authorship and other actions are traceable to the responsible roles, roles were assigned by the designated, authorised roles, and documents and notifications correctly reference other documents according to the process specification.
Since we have exactly one action from a role to a document or notification, we can transform our graph into a tree where each leaf represents a document or notification to be verified (Figure \ref{fig:RSphaseTree}). Edges between the leaves represent references between documents or notifications.
While we distinguish between documents and notifications to match the process
specification, practically, they are only distinct in the kind of data they
refer to. Thus, in our framework we abstract both as simply documents with
different types, as seen in 
Figure~\ref{fig:RSphaseTree}.

\section{Trust on Tracks Framework}

In this section, we describe the setting and parties, then introduce Attestation
Chains as a cryptographic building block that we use together with zk-SNARKs and a standard signature scheme to construct the Trust on Tracks framework.

\subsection{Preliminaries}

For an efficient binary relation $\mathcal{R}$,
a Zero Knowledge Succinct Non-interactive Argument of Knowledge (zk-SNARK) is
the three algorithms $
(\zksnarksetup,
\zksnarkprove,
\zksnarkverify)$
where: 

\begin{description}
\item[
    $\zksnarksetup(1^\lambda,\mathcal{R}) \to \zksnarkcrs$
    ]
    takes a security parameter $\lambda$ and the relation
    $\mathcal{R}$ and outputs a common reference string \zksnarkcrs{}.


    \item[$\zksnarkprove(\zksnarkcrs, \zksnarkpublicin,
    \zksnarkwitness) \to \zksnarkproof$
    ]
    takes input \zksnarkcrs{}, public input \zksnarkpublicin{}, and witness
    \zksnarkwitness{}, and generates the proof \zksnarkproof.

    \item[
    $\zksnarkverify(\zksnarkcrs, \zksnarkpublicin, \zksnarkproof)
    \to \{0, 1\}$
    ]
    outputs $1$ if \zksnarkproof{} is valid.

    \end{description}


We use the notation
$
    \mathcal{R} = \{
        (x; w):
        stmt_1 \land ... \land stmt_n
    \}
$
to describe the relations $\mathcal{R}$ for
zk-SNARK with public inputs $x$, witness $w$, and a set of statements to
prove.

\subsection{Parties \& Setting}

Three kinds of parties participate in our system: One or more
Attestor \attestor{}, a Prover \prover{}, and a Verifier \verifier{}.
An attestor $\attestor$ holds some role $\roleinfo$ and can assign
role $\roleinfo'$ to another attestor $\attestor'$.
Documents are objects created by attestors and represent any kind of data an
attestor may issue.
These can be actual documents, notifications, or arbitrary actions over
other documents.
We refer to a document as \docinfo{}, a collection of documents as a \emph{phase}, and
a group of phases as a \emph{process}.
At the end of a phase, the prover creates a proof which can be confirmed by
anyone acting as a verifier.

\paragraph*{Framework Idea}
Our process begins with a single \emph{root attestor} \rootattestor{} who
\emph{recruits} other attestors by assigning them some \roleinfo{}.
Without loss of generality, we assume that every attestor holds exactly
one \roleinfo{}.
Thus, the group of attestors in a given process naturally map to a tree, with the \emph{root attestor} \rootattestor{} occupying the root, and the
remaining attestors forming the remaining nodes.
Documents issued by an attestor become children of that attestor's node and
their respective branches stop growing, setting them as leaves.
Our goal is to verify the integrity of the tree.
We do this in two steps: (1) we verify that the entire paths from all documents
towards the root are legitimate -- we refer to those paths as \emph{attestation chains}, --
and (2) verify that the attestors and documents conform to business
rules by assigning meaning to all used \docinfo{} and \roleinfo{} values, and
matching them against the process specification.

\subsection{Attestation Chain System}

An \emph{Attestation Chain system} is the tuple of algorithms $(\keygen, \attestdoc,
\attestrole, \verifychain)$ where:

\begin{description}
    \item[$\keygen(\keygeninput) \to (\keypair)$:] Generates key pair.
    \item[$\attestdoc(\attestdocinput) \to \attestdocoutput$:] Outputs document attestation.
     \item[$\attestrole(\attestroleinput) \to \attestroleoutput$:] Outputs role attestation.
      \item[$\verifychain(\verifychaininput)$]
        \hfill \\ 
          $\to \{0, 1\}$:
           Verifies the given attestation chain.
\end{description}

An Attestor \attestor{} generates a dedicated key pair \keypair{} by
running \keygen{}.
\attestor{} is assigned some role \roleinfo{} if they obtain a role
attestation $\rolesig$ over $(\pk, \roleinfo)$ from another attestor
\attestor{}' who runs \attestrole{} with their secret key.
Similarly, an attestor {} may attest to some document information \docinfo{} by
running \attestdoc{} with her secret key.

In our tree model, an Attestation Chain is represented as the sequence of nodes starting with any document
and ending at any attestor on the path to the root attestor.
Thus, an attestation chain is the sequence:
\vspace{-1.6em}
\begin{center}
\footnotesize
$$
(\docinfo, (\pk_0, \roleinfo_0, \docsig),
(\pk_1, \roleinfo_1, \rolesig_0), ..., (\pk_n, \roleinfo_n,
\rolesig_{n-1}))
$$
\end{center}
%
We say that an Attestation Chain is valid iff every attestation in the sequence
(over a document or an attestor) was indeed created by the subsequent attestor,
i.e.,
\begin{align*}
    \attestdoc(\sk_0, \docinfo) &= \docsig  \ \land
    \\
    \attestrole(\sk_1, \roleinfo_0) &= \rolesig_0 \ \land
    \\
    &\vdots
    \\
    \attestrole(\sk_n, \roleinfo_{n-1}) &= \rolesig_{n-1}
\end{align*}
Verifying the tree's integrity thus requires verifying all chains.

\paragraph*{Construction}
We construct the system using a standard signature scheme $\sigscheme =
(\keygen, \sign, \verify)$ where $\keygen(pp)$ generates a key pair
$(\keypair)$, $\sign(\sk, \msg)$ outputs signature $\sig$ for secret key $\sk$
and message $\msg$, and $\verify(\pk, \msg, \sig)$ outputs $1$ if the signature
is valid for $\msg$ under the public key $\pk$ and $0$ otherwise.
Figure~\ref{fig:attchainconstruction} shows the construction.

\begin{figure}
\vspace{0.03in}
\scalecontent{0.917}{%
\subimport{construction/}{layout1}
    }
    \caption{Attestation Chain Construction using Standard Signature Scheme
    \sigscheme{}.}
    \label{fig:attchainconstruction}
\end{figure}

\subsection{The Framework}

We construct a zk-SNARK for verifying process trees using the Attestation
Chain system as a building block.
Given a set of Attestation Chains, a Root Attestor \rootattestor{}, and a
description of business rules, the SNARK verifies all attestation chains with
respect to \rootattestor{} as well as conformance to the business rules.
\subsubsection{Overview}

The SNARK takes as input a root attestor public key $\rpk$, and a set of $\ndocs$
Attestation Chains with size $\chainsize$ each as input.
The root attestor is appointed by the process initiator who thereby requires
all attestation chains to verify with respect to its public key $\rpk{}$.
To perform this verification, the SNARK requires that $\verifychain$ outputs
$1$ for each of the Attestation Chains, and that $\pk_\chainsize = \rpk$
in all of the chains.
This ensures that the final attestation in every chain was created by the root
attestor, effectively setting $\rpk$ at the root of the attestation tree
as visualized in Figure~\ref{fig:fig_snark}.

\tikzstyle{rect} = [
rectangle, minimum width=1cm, minimum height=0.5cm, text
centered, draw=black
]
\tikzstyle{arrow} = [
    thick,
]

\begin{figure}[h] %
\centering
\small
\begin{tikzpicture}
[align=center,node distance=1cm and 0.4cm]
{
\newcommand{\roleinfolabel}[2]{
$\pk_{#1,#2}, \roleinfo_ {#1,#2}, \rolesig_{#1, #2}$
}

\newcommand{\chainnodes}[2]{

\node (docinfo#1) [rect, #2] {$\docinfo_{#1}$};
\node (roleinfo#10) [rect, right=of docinfo#1] {$\pk_{#1, 0}, \docsig_#1$};
\node (roleinfo#1x) [rect, right=of roleinfo#10] {\ldots};
\node (roleinfo#1n) [rect, right=of roleinfo#1x] {$\pk_{#1, n}, \rolesig_{#1, n-1}$};

\draw [arrow] (docinfo#1) -- node[]{} (roleinfo#10);
\draw [arrow] (roleinfo#10) -- node[]{} (roleinfo#1x);
\draw [arrow] (roleinfo#1x) -- node[]{} (roleinfo#1n);
}

%

\chainnodes{0}{}
\chainnodes{1}{below of=docinfo0}
\node (etc) [below of=docinfo1, yshift=0.5cm] {$\vdots$};
\node (etc0) [below of=roleinfo10, yshift=0.5cm] {$\vdots$};
\node (etcx) [below of=roleinfo1x, yshift=0.5cm] {$\vdots$};
\node (etcn) [below of=roleinfo1n, yshift=0.5cm] {$\vdots$};
\chainnodes{l}{below of=etc, yshift=0.2cm}

\node (root) [rect, right=of roleinfo1n] {$\rpk$};

\draw [densely dotted] (roleinfoln.east) -- node[]{} (root.west);
\draw [densely dotted] (roleinfo1n.east) -- node[]{} (root.west);
\draw [densely dotted] (roleinfo0n.east) -- node[]{} (root.west);
\draw [densely dotted] ([xshift=0.8cm,yshift=-0.0cm]etcn.east) -- node[]{} (root.west);

}

\end{tikzpicture}
\caption[Preprocessing of Attestation Chains]{%
    Preprocessing of Attestation Chains.
    All chains are extended with the root attestor public key $\rpk$, 
    forming a tree of attestations rooted at $\rpk$.
    }
    \label{fig:fig_snark}
    \end{figure}
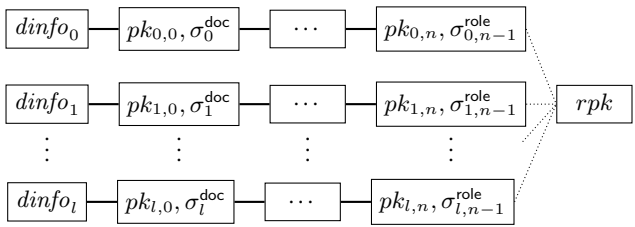

\paragraph*{Business Rules}
In addition to the cryptographic validity of the attestation chains, the SNARK
enables two types of functional verifications to be specifiable
via a Chain Policy function $\chainpolicy(\docinfo, \{\roleinfo_i\}^\chainsize)$ $ \to \{0,
1\}$, and a Phase Policy function $\phasepolicy(\{\docinfo_i\}^\ndocs) \to \{0,
1\}$.
The Chain Policy takes as input the set of all $\docinfo, \roleinfo$ building a
particular attestation chain of size $n$.
Within the function, meaning can be assigned to the raw values of $\roleinfo$
and get checked individually or in relation to any other given $\roleinfo' \ne
\roleinfo$.
For example, one can model an attribute-based access control by encoding
attribute names into $\roleinfo$ and checking in $\chainpolicy$ whether
attributes of a given $\roleinfo$ allow creating $\docinfo$.

Analogously, the Phase Policy function \phasepolicy{} verifies the meaning
assigned to all $\docinfo$ across the entire set of all attestation chains
as well as associations across them.
For example, given $(\docinfo_1, \docinfo_2)$ as inputs to $\phasepolicy$, one
can require that $\docinfo_1$ is a review report for $\docinfo_2$.

\subsubsection{zk-SNARK Definition}

We require a zk-SNARK that proves the following relation:
\begin{align*}
    &R = 
    \big\{
        (rpk, \phasepolicy, \chainpolicy);
        \{
            (
                \docinfo_i,
                \pk_{i, 0},
                \roleinfo_{i, 0},
                \docsig_i, \\
                &
                \ \ \ \ \ \ \ \ 
                \ \ \ \ \ \ \ \ 
                \ \ \ \ \ \ \ \ 
                \ \ \ \ \ \ \ \ 
                \{
                    (
                        \pk_{i,j},
                        \roleinfo_{i,j},
                        \rolesig_{i, j}
                    )
                \}^{\chainsize_i}
            )
        \}^\ndocs
    \big\}:
    \\
    &
    \ \ \ \ 
    \ \ \ \ 
    \ \ \ 
    \phasepolicy(\docinfo_0, ..., \docinfo_\ndocs) \ \ \ = 1
    \ \land
    \\
    &
    \forall {i \in [\ndocs]}\ \ 
    \chainpolicy(\roleinfo_{i, 0}, ..., \roleinfo_{i, \chainsize_i}) = 1
    \ \land \\
    &
    \ \ \ \ \ 
    \ \ \ \ \  \ \ 
    \ \ \ \ \  \ \ 
    \ \ \ \ \  \ \ 
    \ \ \ \  \ \ 
    \rpk = \pk_{i, \chainsize_i} \ \land \\
    &
    \ \ \ \ \ 
    \ \ \ \ \  \ \ 
    \verifychain(
                \docinfo_i,
                \pk_{i, 0},
                \roleinfo_{i, 0},
                \docsig_i, \\
                &
                \ \ \ \ \ 
                \ \ \ \ \ \ 
                \ \ \ \ \ \ 
                \ \ \ \ \ \ 
                \ \ \ \ \ \  \ \ 
                \{
                    (
                        \pk_{i,j},
                        \roleinfo_{i,j},
                        \rolesig_{i, j}
                    )
                \}^{\chainsize_i} 
    ) = 1
\end{align*}

\section{Evaluation}

In this section, we evaluate our Trust on Tracks framework by instantiating it
for the first phase of Sektorleitlinie 22, the requirements specification phase, as described in Section~\ref{section:railway}.

\subsection{Railway Instance}

To instantiate our framework for Phase 1 of Sektorleitlinie 22, we implement
the Attestation Chain Scheme, the zk-SNARK, as well the chain and phase policy
verification functions.
For constructing zk-SNARK circuits, high-level languages, such as
ZoKrates~\cite{zokrates}, provide a syntax that closely resemble familiar programming
languages, abstracting away the underlying circuit details and complexities.
To be compatible with ZoKrates, we realise our Attestation Chain scheme using
EdDSA over the BabyJubJub curve, which corresponds to the scalar field of BN128
which is used by ZoKrates.
The code is open source and available on Github~\footnote{https://github.com/tgalal/trust-on-tracks}.
\\

\subsubsection{Modelling \docinfo{}}

We encode into \docinfo{} the information
$(\mathtt{doctype},$ $ \mathtt{identifier}$, 
$\mathtt{ref})$
where
$\mathtt{doctype}$ denotes the type of a document,
$\mathtt{identifier}$ is a SHA-256 hash of the document, and $\mathtt{ref}$ is
a SHA-256 hash of concatenated $\mathtt{identifier}$  values of any referenced
\docinfo{}.

Phase 1, as described in Section~\ref{section:railway}, requires
\sktSA{},
\sktTR{},
\sktPTD{},
\sktPTDD{},
\sktRS{},
\sktSUC{},
\sktSUCC{},
\sktFGV{},
\sktFGVV{}
as deliverable document types $\mathtt{doctype}$.
The $\mathtt{identifier}$ field is populated by the SHA-256 hash of
the digital copy of the respective document.
Finally, according to the specification, \sktPTD{} references
\sktRS{},
\sktPTDD{},
\sktTR{}, 
and \sktSA{},
while \sktSUC{} references \sktSUCC{}.
To model those relations, we generate $\mathtt{ref}$ for \sktPTD{} as 
SHA-256(%
$
\mathtt{ID(\sktRS)} \|
\mathtt{ID(\sktPTDD)} \|
\mathtt{ID(\sktTR)} \|
\mathtt{ID(\sktSA)}
$%
)
and for \sktSUC{} as 
SHA-256(%
$
\mathtt{ID(\sktSUCC)} 
$%
)
where $\mathtt{ID(doctype)}$ outputs the document identifier for a 
document of type $\mathtt{doctype}$.

\subsubsection{Modelling \roleinfo{}}

We encode into \roleinfo{} the set of document types an Attestor may create.
This is a subset of the power set
$\mathcal{P}(\{\sktSA, \sktTR, \sktPTD, \sktRS,$ $\sktSUC, \sktPTDD, \sktSUC2,
\sktFGV, \sktFGVV\})$.
We implement those sets by encoding every $\mathtt{doctype}$ as a unique power of
$2$ integer, and setting the value of \roleinfo{} to the bit-wise OR of all
permitted document types.

\subsubsection{Chain \& Phase Policies}
In the Chain Policy function $\chainpolicy$ we implement hierarchical
attribute based roles that support delegation of attributes down the
hierarchy.
Attributes in $\roleinfo$ represent document types the holder is
allowed to create or delegate a subset of to another attestor.
This places every Attestor as an \emph{Attribute Authority} for the attributes
it holds.

In the Phase Policy function $\phasepolicy$ 
we implement the business rules for the first phase \sekt{}.
Therein, the function verifies the references between the submitted documents.

\subsection{Performance}

\begin{table*}
    \centering
    \caption{
        Execution Time and Memory Usage for various exemplary inputs and
        the \sekt{} specification.
     }
    \label{table:performance}
{%
\newcommand{\theaderr}[1]{\textbf{#1}}
\newcommand{\theader}[1]{\footnotesize #1}
\begin{tabular}[]{c c c c || c c || c c || c c}
\toprule
\multicolumn{4}{c ||}{\theader{Count}} &
\multicolumn{2}{c ||}{\theader{Compile}} &
\multicolumn{2}{c ||}{\theader{Setup}} &
\multicolumn{2}{c}{\theader{Proof Generation}}
\\
\theaderr{Documents} & \theaderr{Attestations} & \theaderr{Signatures} & \theaderr{Constraints} &
\theaderr{Time} & \theaderr{Memory} &
\theaderr{Time} & \theaderr{Memory} &
\theaderr{Time} & \theaderr{Memory}
\\
\midrule
\multicolumn{10}{c}{Benchmarks with Exemplary Inputs} \\
\midrule
1 & 1 & 1 & 101775 &
7s & 0.9 GB &
4s & 0.4 GB &
6s & 0.5 GB
\\
1 & 2 & 2 & 203314 &
14s & 1.9 GB &
8s & 0.9 GB &
12s & 1 GB
\\
2 & 2 & 4 & 406628 &
30s & 3.8 GB &
15s & 1.8 GB &
24s & 2 GB
\\
\midrule
\multicolumn{10}{c}{Phase 1 of \sekt} \\
\midrule
9 & 3 & 27 & 2843728 &
266s & 26.1 GB &
118s & 12.3GB &
167s & 13.6GB
\\
\bottomrule
\end{tabular}
}

\end{table*}

We execute our implementation on a \texttt{c5.4xlarge} EC2 instance with 32 GB
memory and 16 3.00 GHz vCPUs.
We measure the execution time and maximum memory consumption for compilation,
setup, and proof generation for different attestation tree sizes and show the
results in 
Table~\ref{table:performance}, divided in two sections.
The upper section shows measurements for arbitrary input sizes and shows that
the execution times and memory usage grow linear in the total number of
signatures.
The total number of signatures is influenced by the number of input documents
together with the tree depth, which in turn influences the number of
attestations per document.
To simplify the implementation, we fix the path from any document to the root, i.e., the number of attestations per document, to always match the tree depth.
Thus, the total number of signatures is always equal the number of documents $\times$
tree depth.
The lower section of the table contains measurements obtained from the
configuration of Phase 1 of \sekt{} specification.
The number of CPUs does not impact the performance as all process steps, i.e. compilation, setup, and proof generation, are single-threaded.

\subsection{Publishing Proofs on the Blockchain}

Proofs generated by ZoKrates are highly optimized for storage and verification
on the Ethereum blockchain.
In fact, ZoKrates generates the respective smart contract code in solidity,
ready for deployment on the blockchain.
Using the blockchain for storage and verification of those proofs strengthens
trust in the business process.
On the one hand, the blockchain acts as an immutable storage for those proofs,
on the other hand, private auditors can verify the consistency between
published proofs and any confidential data used in creating them - as part
of any necessary auditing procedure.

It costs $5193982$ Gas to deploy the ZoKrates-generated smart contract for
Trust on Tracks, and $463359$ Gas to store a single proof on the blockchain.
On-chain verification of a proof costs $3679995$ Gas, expected to be paid once
by the prover.
For an exemplary gas price of 0.42 gwei and Ethereum price of 1800 USD, these
gas costs approximately correspond to \$4.67, \$0.42, \$3.31 respectively.

The prover may also choose to split a process into phases and generate a proof
for every phase individually.
This has the advantage of added granularity into the progress of
a process that gets communicated to verifiers.
In this case, the costs for storing and verifying the different proofs are
summed, but deploying the contract is largely not affected.

\section{Discussion}

\subsection{Trust on Tracks vs Blockchain-only}

Modelling an entire process on blockchain \cite{FRUU18,DSLT20,Bau24} has
drawbacks.
It either requires the migration of an existing and working process,
to be coordinated and executed entirely on the blockchain in form of a smart
contract, or running the smart contract in sync with the 
process, mirroring every action or step taken.
The former is unlikely to be adopted, and the latter adds complexity.
Furthermore, dealing with confidential data presents a challenge to a
blockchain-only solution and typically invokes off-chain workarounds such as
IPFS.

We identified public transparency as a main motivation behind these works
i.e., proving that a process was correctly executed.
Trust on Tracks achieves that without modelling the entire process
using smart contracts, and inherently enables confidentiality for any aspect of
the process.

\subsection{Tree Optimization}

There are a few optimizable aspects in Trust on Tracks:

\subsubsection{Redundant Signatures}
\label{section:redundantsigs}
All the attestation chains of a particular tree are of the same size that
corresponds to the tree height; this carries a performance penalty.
More specifically, according to Figure~\ref{fig:RSphaseTree}, there should be
exactly 17 unique digital signatures in total, however our evaluation
produced 27 for the same process.
This discrepancy is attributed to a trade-off between
flexibility and efficiency in modelling a tree using ZoKrates.
When the circuit is compiled, all input sizes must be fixed.
This would require determining the exact structure of the tree beforehand, i.e.,
the sizes of all branches and how the nodes are connected.
Instead, we let the prover decide this structure and it gets checked during
verification.
Only the tree height is fixed during compilation, which resembles the maximum
size of an attestation chain.
Thus in Figure~\ref{fig:RSphaseTree}, although the chain from FRA to SA should
optimally be of size $2$, it is of size $3$.
As the overall complexity of a ZoKrates program is the sum of the cost of
\emph{all branches}, it does not matter whether this extra attestation is
skipped or verified, hence we counted it as a redundant
signature.

\subsubsection{Redundant Branches}

Additional redundant signatures appear due to representing the tree as paths
from leaves to root.
According to Figure~\ref{fig:RSphaseTree}, the Operator subtree should 
have 8 signatures.
In reality, processing this subtree results in 10 signatures.
Although the same BAV node lies on two attestation chains: from
Operator to RS, and from Operator to SUC, the Operator's signature over BAV is
verified twice.
This is because we do not cross-check if an attestation that exists on two
paths has been already verified and can be skipped.
However, due to the reasons outlined in Section~\ref{section:redundantsigs}, simply
cross-checking will not have a performance benefit as all branches in the
program are executed.
Instead, addressing this requires an alternative and more complex
tree representation.

\vspace{-0.05em}
\section{Conclusion}

We presented the Trust on Tracks framework for adding public transparency to
processes, inspired by the intricacies of the German guidelines for railway
modifications \sekt{}.
In contrast to other approaches, our framework does not require 
existing processes to be modified or executed on the blockchain, instead, processes
still execute off-chain, and only proofs of correctness are published
on-chain for public verifiability and transparency.
We evaluated the framework by implementing it for the first Phase of German
railway guidelines and have shown its practicality for adoption.

\section*{Author Statement}
Valeria Tisch developed our blockchain-only solution and manuscript, which Katja Assaf and Andreas Polze supervised.
Tarek Galal and Katja Assaf developed the Trust on Tracks solution and wrote the manuscript.
Tarek Galal provided the implementation.

\bibliographystyle{styles/ieee/IEEEtran}
\bibliography{%
    bib/railway,
    bib/other.bib,
    bib/blockchain_only.bib,
    bib/zk_bpm.bib
}

\end{document}